\documentclass[leqno]{article}
\usepackage{mte06e}
\usepackage{amsmath}
\usepackage{amssymb}
\usepackage{amsfonts}
\usepackage[mathscr]{eucal}

\usepackage{graphicx}% Include figure files

\newdefin{dfn}{Definition}

\newdefin{remark}{Remark}

\begin{document}
\keywords{Phase transition, Surface model, Curvature Hamiltonian}
%\mathclass{Primary **C**; Secondary **D**.}
%\keywords{These must be included}

\abbrevauthors{T. Endo et. al.}
\abbrevtitle{Phase transition on a disk}

\title{Phase transition of \\extrinsic curvature surface model on a disk}

\author{T. Endo, M. Egashira, S. Obata, H. Koibuchi}
\address{Department of Mechanical and Systems Engineering, Ibaraki National College of Technology\\
Nakane 866, Hitachinaka-shi, Ibaraki 312-8508, Japan\\
E-mail: koibuchi@mech.ibaraki-ct.ac.jp}
\thanks{This work is supported in part by a Grant-in-Aid for Scientific Research, No. 18560185.}

\maketitleMTE

\begin{abstract}
An extrinsic curvature surface model is investigated by Monte Carlo simulations on a disk. We found that the model undergoes a first-order transition separating the smooth phase from the collapsed phase. The results in this paper together with the previous ones suggest that the order of the transition is independent of whether the surface is compact (closed) or non-compact (open). 
\end{abstract}

%----------------------------------------------------------
\section{Introduction}\label{intro}
%----------------------------------------------------------
This paper is aimed at showing that a surface model undergoes a first-order transition, which separates the smooth phase from the collapsed phase on a disk with an extrinsic curvature. The results in this paper together with previous ones \cite{Kownacki-Diep-PRE-2002,KOIB-PRE-2004-1,KOIB-PRE-2005,KOIB-NPB-2006,KOIB-PLA-2006-2} indicate that the first-order transition occurs independent of whether the surface is closed or open. 

Two-dimensional surface emerges as an interface between two different materials including the air. Biological membranes are considered to be a two-dimensional surface separating two-different biological materials \cite{NELSON-SMMS-2004,Gompper-Schick-PTC-1994,BOWICK-TRAVESSET-PREP-2001,WIESE-PTCP19-2000,WHEATER-JP-1994}, and they are described by some curvature Hamiltonian such as the one of Helfrich, Polyakov and Kleinert \cite{HELFRICH-NF-1973,POLYAKOV-NPB-1986,Kleinert-PLB-1986}. 

Fluctuations of surfaces are considered to soften/stiffen the surface \cite{NELSON-SMMS-2004-2}, and theoretical investigations \cite{DavidGuitter-EPL-1988,Peliti-Leibler-PRL-1985,BKS-PLA-2000,BK-PRB-2001,Kleinert-EPJB-1999} as well as numerical ones \cite{BCFTA-1996-1997,WHEATER-NPB-1996,KANTOR-NELSON-PRA-1987,GOMPPER-KROLL-PRE-1995,KOIB-PRE-2003,KOIB-PRE-2004-2,KOIB-PLA-2003-2,KOIB-EPJB-2004,KOIB-PLA-2005-1,KOIB-PLA-2006-1,KANTOR-SMMS-2004,KANTOR-KARDER-NELSON-PRA-1987,BCTT-EPJE-2001,BOWICK-SMMS-2004} have been made on this problem. Surface models are classified into two classes: one is the extrinsic curvature model and the other is the intrinsic curvature model. It was reported that the intrinsic curvature models undergo a first-order transition \cite{KOIB-EPJB-2004,KOIB-PLA-2006-1,KOIB-PLA-2005-1}, and we know that the transition is independent of whether the surface is closed \cite{KOIB-EPJB-2004,KOIB-PLA-2006-1} or open \cite{KOIB-PLA-2005-1}. 

The surface model with extrinsic curvature is also known to have a long-range order although the surface is softened by surface fluctuations. It was reported that the extrinsic curvature model undergoes a first-order transition on triangulated spheres \cite{Kownacki-Diep-PRE-2002,KOIB-PRE-2004-1,KOIB-PRE-2005,KOIB-NPB-2006} as well as on triangulated tori \cite{KOIB-PLA-2006-2}. 

However, it is still unknown whether the transition occurs on a disk with extrinsic curvatures. Many biological membranes are closed; the membranes play a role of an interface between two different materials, however, we know that there are holes in biological membranes. We must recall that a membrane with a hole is topologically equivalent to the disk. World surfaces of the string model can be divided not only into closed surfaces but also into open surfaces \cite{WHEATER-JP-1994}. Moreover, it is still unclear how the open boundary influences the transition of the extrinsic curvature models.

Therefore, it is interesting to study the phase structure of the surface model on a triangulated disk. In order to clarify this point, we study the surface model by using the Monte Carlo (MC) simulations; the model is defined by a Hamiltonian, which is a linear combination of the Gaussian bond potential and a bending energy. 

%------------------------------------------
\section{Model}
%------------------------------------------
%++++++++++++++++++++++++++++++++++
\begin{figure}[hbt]
\centering
\includegraphics[width=4.5cm,height=4.1cm]{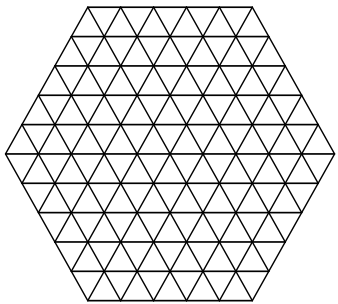}
\caption{A triangulated disk of size $(N,N_E,N_P)\!=\!(91,240,150)$, where $N,N_E,N_P$ are the total number of vertices, the total number of edges, the total number of plaquettes (or triangles).}
\label{fig-1}
\end{figure}
%++++++++++++++++++++++++++++++++++
Figure \ref{fig-1} is a triangulated disk of size $N\!=\!91$, which is obtained by dividing each edge of the original hexagon into $5$-pieces. By dividing each edge into $L$-pieces we have a lattice of $(N,N_E,N_P)\!=\!(3L^2\!+\!3L\!+\!1, 9L^2\!+\!3L,6L^2)$, where $N,N_E,N_P$ are the total number of vertices, the total number of edges, and the total number of plaquettes (or triangles). The flat and regularly triangulated disk such as the one shown in Fig.\ref{fig-1} is used as a starting configuration of MC simulations.

As stated at the end of the introduction we investigate the phase structure of an extrinsic curvature surface model on the triangulated disk. The model is defined by Hamiltonian $S$ that is the linear combination of the Gaussian bond potential $S_1$ and the extrinsic curvature energy $S_2$:
\begin{eqnarray}
\label{S1S2}
&&S=\lambda S_1+bS_2, \quad S_1=\sum_{(ij)} \left(X_i-X_j\right)^2,\\
&&S_2=\sum_i\sum_{j(i)}\left[1-{\bf n}(i)\cdot {\bf n}_{j(i)}\right], \nonumber
\end{eqnarray}
where $\sum_{(ij)}$ in $S_1$ is the sum over bond $(ij)$ connecting the vertices $X_i \in {\bf R}^3$ and $X_j\in {\bf R}^3$, and $\lambda$ is the surface tension coefficient, which is assumed to  $\lambda\!=\!1$, and $b$ is the bending rigidity. The vector ${\bf n}(i)$ in $S_2$ is the unit normal vector at the vertex $i$ and is defined by 
\begin{equation}
\label{normal}
{\bf n}(i)= {{\bf N}_i \over \vert {\bf N}_i\vert}, \quad {\bf N}_i = \sum _{j(i)} {\bf  n}_{j(i)} A_{\it \Delta_{j(i)}},
\end{equation}
where $\sum _{j(i)}$ denotes the sum over triangles  $j(i)$ linked to the vertex $i$. The vector ${\bf n}_{j(i)}$ is the unit normal of the triangle $j(i)$, and  $A_{\it \Delta_{j(i)}}$ is the area of the triangle $j(i)$.  We note that $\sum_{(ij)}$ in $S_1$ includes the sum over the boundary bonds. As a consequence, the length of the boundary bonds can also vary in the simulations. Moreover, the boundary vertices share the bending energy $S_2$ in Eq.(\ref{S1S2}), because $S_2$ is defined not only on the internal vertices but also on the boundary vertices.

\begin{figure}[hbt]
\begin{picture}( 10,10)(  0,0)
\put(60,-10){\makebox(0,0){(a) The model in this paper}}%
\put(210,-10){\makebox(0,0){(b) The conventional model}}%
\end{picture}%
\vspace{0.5cm}
\centering
\includegraphics[width=9cm]{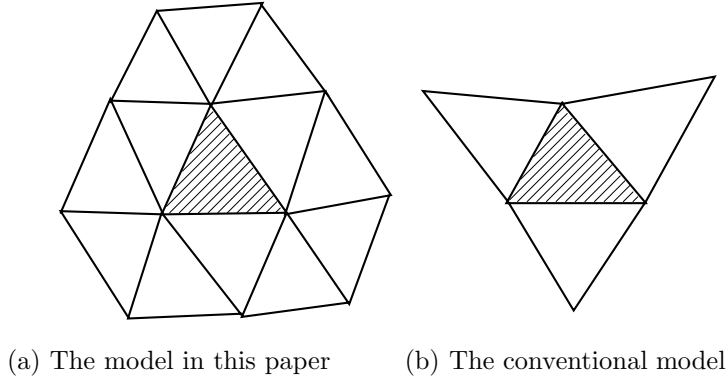}
\caption{The range of the interaction between the normal vectors in (a) the model of Eq.(\ref{S1S2}) and in (b) the conventional model. The normal of the shaded triangle interacts with the normals of the other triangles shown in the figure.}
\label{fig-2}
\end{figure}

In order to understand the interaction of $S_2$ in Eq. (\ref{S1S2}), we graphically show in Fig. \ref{fig-2}(a) the interaction range between the normals of triangles.  Figure \ref{fig-2}(b) is the interaction range defined by the conventional bending energy of the type $1-{\bf n}_i\cdot {\bf n}_j$, where ${\bf n}_i$ is the unit normal vector of the triangle $i$. The normal of the shaded triangle interacts with the normals of the other triangles in the figure. The range of interactions in Fig. \ref{fig-2}(a) is relatively larger than that in Fig. \ref{fig-2}(b).
%++++++++++++++++++++++++++++++++++
%++++++++++++++++++++++++++++++++++

The partition function is defined by 
\begin{equation}
 \label{partition-function}
Z(b) = \int \prod _{i=1}^N dX_i \exp\left[-S(X)\right], \quad S(X)=S_1+bS_2,
\end{equation}
where $N$ is the total number of vertices as described above, and $\prod _{i=1}^N dX_i$ is a $3N$-dimensional integration. The expression $S(X)$  shows that $S$ explicitly depends on the variable $X=(X_1,X_2,\cdots, X_N)$, where $X_i \in {\bf R}^3$ is the position of the vertex $i$ as described above. The center of the surface is fixed in the integration to remove the translational zero-mode.

Note that the surface can be understood as the image of a mapping $X$ from a two-dimensional parameter space $P$ such as shown in Fig.\ref{fig-1} to ${\bf R}^3$; $X : P \to  {\bf R}^3$. The surface is allowed to self-intersect in our model and, hence it is called a {\it phantom surface}. This implies that the mapping $X$ is not always injective.

A physical observable $\langle Q \rangle$ is obtained by averaging the quantity $Q(X)$ with the partition function such that
\begin{equation}
 \label{phsyical-quantity}
\langle Q \rangle =  {1\over  Z(b)}\int \prod _{i=1}^N dX_i Q(X) \exp\left[-S(X)\right].
\end{equation}

For $Q=S_2$, we have the expression
\begin{equation}
 \label{bending-energy}
\langle S_2 \rangle =  -{\partial \log Z \over \partial b} = -{1\over Z}{\partial Z \over \partial b},
\end{equation}
where $Z$ is dependent on the bending rigidity $b$ and the surface tension $\lambda$, which is assumed to be $\lambda\!=\!1$ as described above. The specific heat $C_{S_2}$ can also be defined by a second-derivative of $Z$ such that
\begin{equation}
 \label{specific-heat}
\langle C_{S_2} \rangle = -{b^2 \over N} {\partial  \langle S_2 \rangle \over \partial b}= {b^2 \over N}{\partial^2 \log Z \over \partial^2 b}.
\end{equation}

A phase transition is called {\it first-order} or {\it discontinuous} if the first-derivative of $Z$ with respect to a parameter is a discontinuous function. Moreover, a phase transition is called {\it second-order} or {\it continuous} if the first-derivative of $Z$ with respect to a parameter is continuous and the second-derivative of $Z$ is discontinuous. Thus, the order of phase transition is originally connected to the analytic property of $Z$. 

We have another definition on the order of phase transitions. A phase transition can be called discontinuous if there is a physical quantity that discontinuously changes against a given parameter such as the temperature (or the bending rigidity in the surface model). 

 Therefore, we can call the phase transition of the model as a first-order one not only if $S_2$ is discontinuous against $b$ but also if some physical quantity is discontinuous against $b$. 

%------------------------------------------
\section{Monte Carlo technique}\label{MC-Techniques}
%------------------------------------------
The $3N$-dimensional integration $\prod _{i=1}^N dX_i$ in $Z$ can be performed numerically. The canonical Monte Carlo technique is used to update the variables $X (\in {\bf R}^3)$ so that $X^\prime \!=\! X \!+\! \delta X$, where the small change $\delta X$ is made at random in a small sphere in ${\bf R}^3$. The new position $X^\prime$ is accepted with the probability ${\rm Min}[1,\exp(-\Delta S)]$, where $\Delta S\!=\! S({\rm new})\!-\!S({\rm old})$. The radius $\delta r$ of the small sphere is fixed at the beginning of the simulations to maintain about $50\%$ acceptance rate.  This procedure is called the Metropolis MC algorithm, and it produces equilibrium surface configurations $X$ which satisfy the Boltzmann distributuion function $\exp [-S(X)]$.

We use a random number called Mersenne Twister \cite{Matsumoto-Nishimura-1998} in the MC simulations. A sequence of random number is used for the 3-dimensional move of vertices $X$ and  for the Metropolis accept/reject of the update of $X$. 

By using a sequence of the equilibrium configurations of surfaces $\{X(i)\}$ produced by the simulations, we have $\langle Q \rangle$ in Eq.(\ref{phsyical-quantity}) simply by
\begin{equation}
 \label{numerical-average}
\langle Q \rangle =  {\sum_i Q[X(i)] \over \sum_i 1}.
\end{equation}
In the following, we write $\langle Q \rangle$ as $Q$ for simplicity.

 We use the surfaces of size $N\!=\!25117$, $N\!=\!15337$, $N\!=\!9577$, $N\!=\!5167$, $N\!=\!2977$,  and $N\!=\!1387$.  The size of the surfaces is obviously larger than that of the same model on spherical surfaces used in \cite{KOIB-PRE-2004-1,KOIB-PRE-2005,KOIB-NPB-2006}.

The total number of Monte Carlo sweeps (MCS) after the thermalization MCS is about $4.7\!\times\!10^8$ close to the transition point of surfaces of size $N\!=\!25117$ and $N\!=\!15337$, $3.0\!\times\!10^8\sim 2.6\!\times\!10^8$ for $N\!=\!9577$ and $N\!=\!5167$, and $1.5\!\times\!10^8\sim 1\!\times\!10^8$ for  $N\!=\!2977$ and $N\!=\!1387$. Relatively smaller number of MCS is done at non-transition points in each surface. 

%------------------------------------------
\section{Simulation Results}\label{Results}
%------------------------------------------
%++++++++++++++++++++++++++++++++++
\begin{figure}[hbt]
%\vspace{0.5cm}
%\centering
\unitlength 0.1in
\begin{picture}( 10,10)(  0,0)
\put(17,34){\makebox(0,0){(a) $b\!=\!0.458$ (collapsed)}}%
\put(32,34){\makebox(0,0){(b) $b\!=\!0.458$ (smooth)}}%
\put(17,-1.5){\makebox(0,0){(c) The surface section}}%
\put(33,-1.5){\makebox(0,0){(d) The surface section}}%
\end{picture}%
\vspace{0.5cm}
\includegraphics[width=7.5cm]{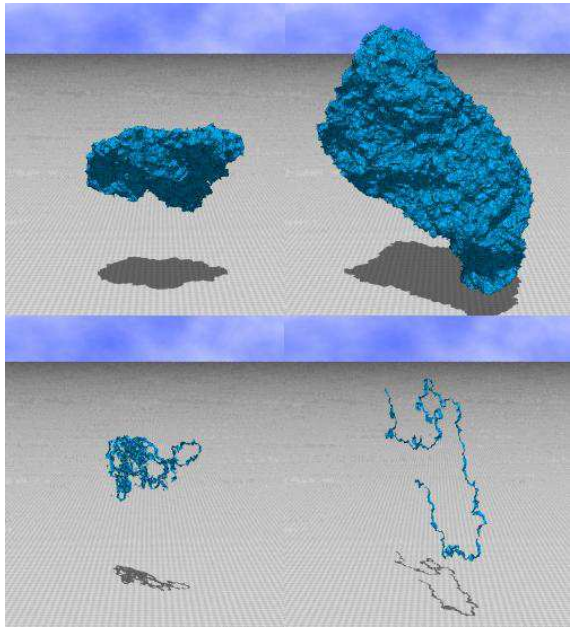}
\caption{ Snapshots of the surface of $N\!=\!25117$ obtained at $b\!=\!0.458$; (a) a collapsed surface, (b) a smooth surface,  (c) the surface section of (a), and (d) the surface section of (b). }
\label{fig-3}
\end{figure}
%++++++++++++++++++++++++++++++++++
First, we show snapshots of surfaces of size $N\!=\!25117$ in Figs.\ref{fig-3}(a), and \ref{fig-3}(b) obtained at $b\!=\!0.458$. The surface sections of them are shown in Figs.\ref{fig-3}(c), and \ref{fig-3}(d). The surface in Fig.\ref{fig-3}(a) is a collapsed surface, and that in  Fig.\ref{fig-3}(b) is a smooth one.

%++++++++++++++++++++++++++++++++++
\begin{figure}[hbt]
\centering
\includegraphics[width=9.0cm]{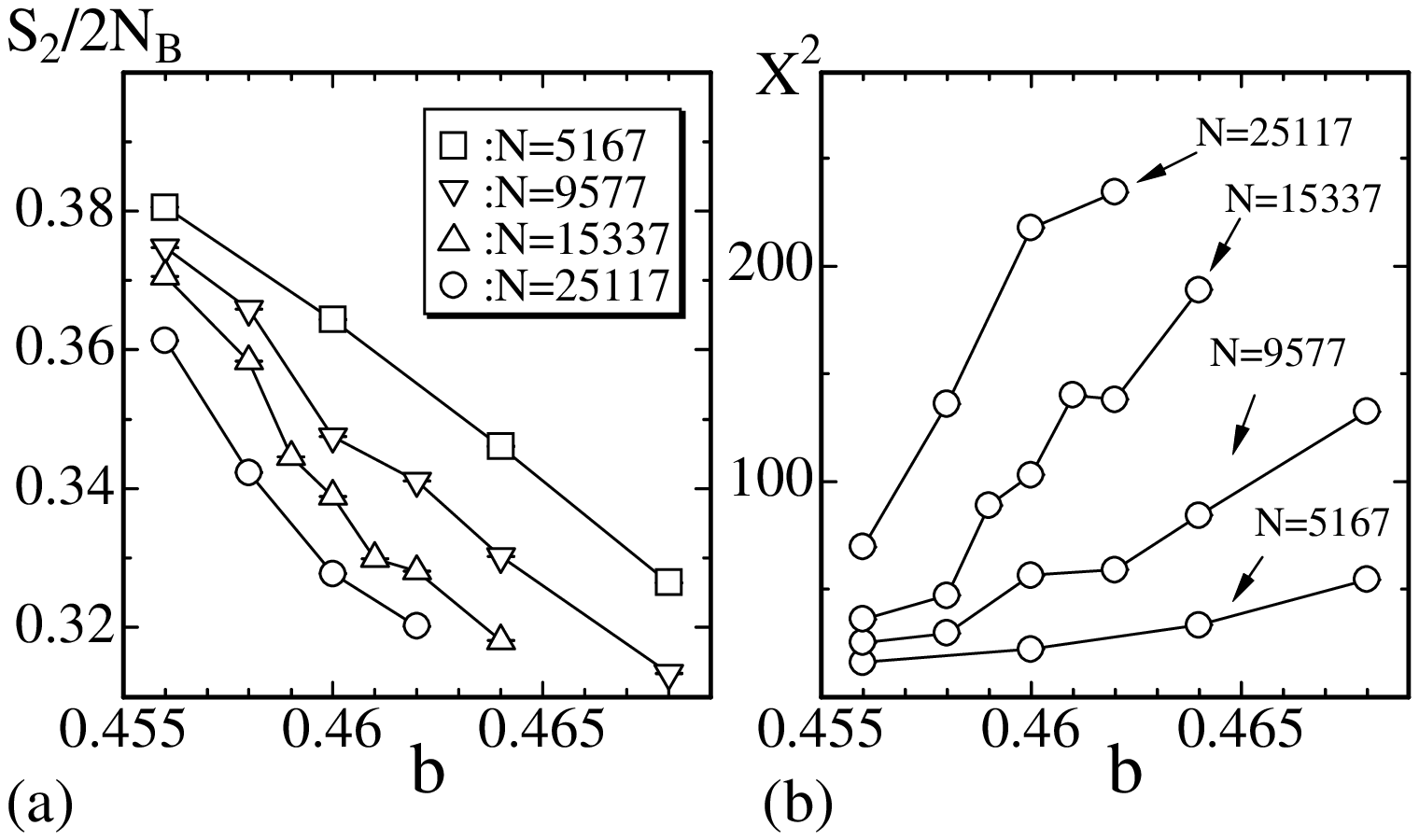}
\caption{(a) The bending energy $S_2/2N_B$ vs. $b$, and (b) the mean square size is $X^2$ vs. $b$, obtained on surfaces of $N\!=\!25117\sim N\!=\!5167$.}
\label{fig-4}
\end{figure}
%++++++++++++++++++++++++++++++++++
Figure \ref{fig-4}(a) shows the bending energy $S_2/2N_B$ against $b$, where $N_B$ is the total number of bonds including the boundary bonds. Dividing $S_2$ by $2N_B$ we have the bending energy per bond, where $2N_B$ comes from the definition of $S_2$ in Eq.(\ref{S1S2}). We find that $S_2/2N_B$ rapidly changes against $b$ when $N$ becomes large. This indicates that a phase transition occurs between the smooth phase and some non-smooth phase, although no discontinuous change can be seen in $S_2/2N_B$. 
The reason why we can see no discontinuity in $S_2/2N_B$ is because of the finiteness of $N$ in the numerical simulations. 

The mean square size $X^2$ is defined by
\begin{equation}
\label{X2}
X^2={1\over N} \sum_i \left(X_i-\bar X\right)^2, \quad \bar X={1\over N} \sum_i X_i,
\end{equation}
where $\bar X$ is the center of the surface, and $X^2$ are plotted in Fig.\ref{fig-4}(b). No discontinuous change can be seen in $X^2$ in the figure, however, we find that the surface size rapidly changes as $N$ increases. This reflects that the surface size discontinuously changes at the transition point in the limit of $N\to \infty$; the model undergoes the collapsing transition.

%++++++++++++++++++++++++++++++++++
\begin{figure}[hbt]
\centering
\includegraphics[width=9.0cm]{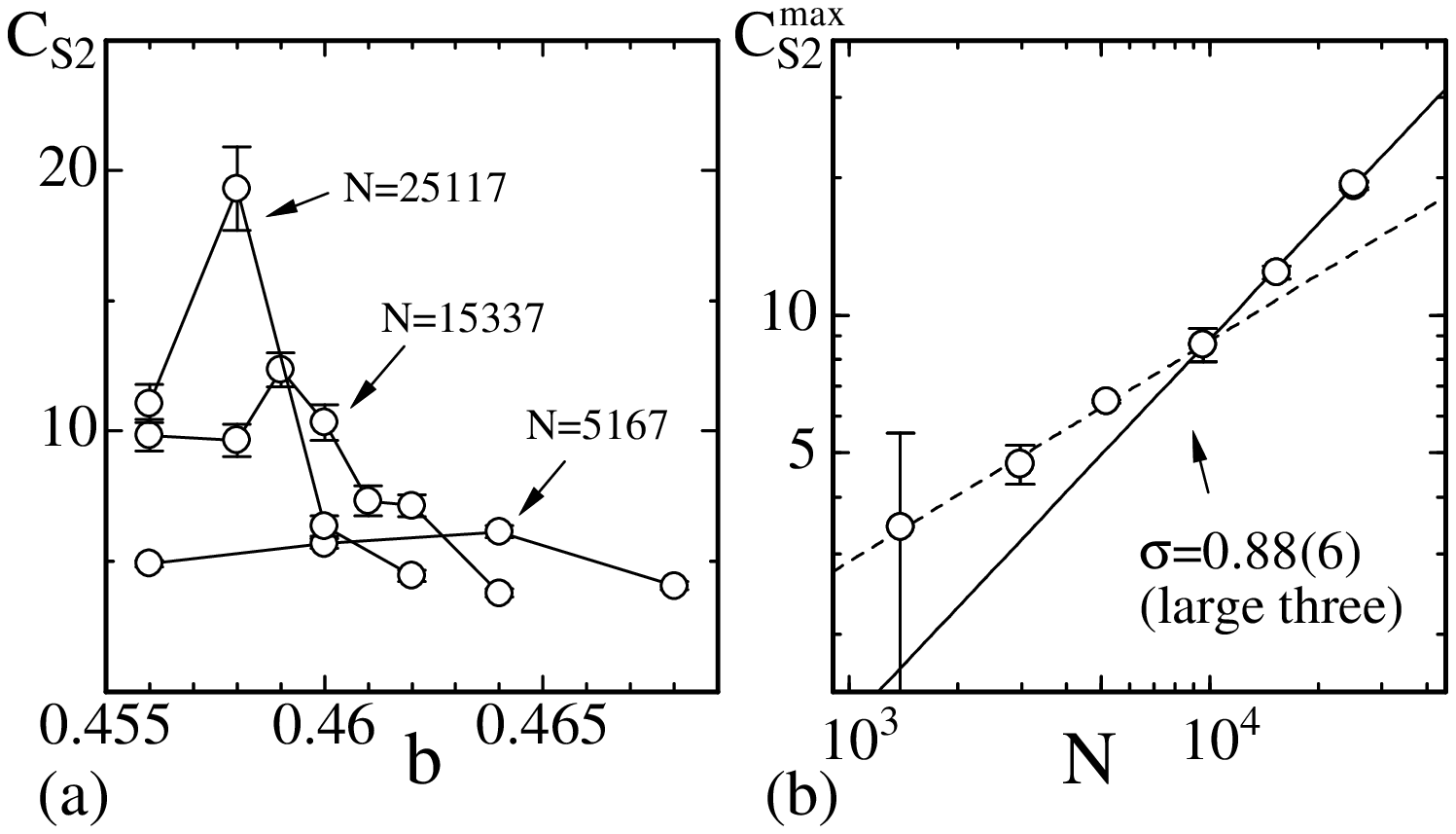}
\caption{(a) The specific heat $C_{S_2}$ vs. $b$, and (b) the peak values $C_{S_2}^{\rm max}$ vs. $N$ in a log-log scale. The solid line is drawn by fitting the largest three data to Eq.(\ref{sigma-fitting}) and indicate a discontinuous transition. The dashed line is drawn by fitting the smallest four data to Eq.(\ref{sigma-fitting}) and is consistent with a continuous transition when $N\!\leq\!9577$. The error bars denote the statistical errors. }
\label{fig-5}
\end{figure}
%++++++++++++++++++++++++++++++++++
The specific heat $C_{S_2}$ of the bending energy $S_2$ can also be written as
\begin{equation}
\label{Spec-Heat}
C_{S_2} = {b^2\over N} \langle \; \left( S_2 - \langle S_2 \rangle\right)^2 \; \rangle, 
\end{equation}
and are shown in Fig. \ref{fig-5}(a) against $b$. The errors shown as error bars in the figure are the statistical ones, which were obtained by the standard binning analysys. Anomalous behavior is  obviously seen in $C_{S_2}$ at each $N$, and it indicates the existence of the surface fluctuation transition, since the anomalous peak of $C_{S_2}$ implies that the fluctuation of the surface becomes very large at the peak point.   

We plot the peak value $C_{S_2}^{\rm max}$ against $N$ in Fig.\ref{fig-5}(b) in a log-log scale in order to clarify the order of the transition. The solid straight line in Fig.\ref{fig-5}(b) was drawn by fitting the largest three $C_{S_2}^{\rm max}$ to 
\begin{equation}
\label{sigma-fitting}
C_{S_2}^{\rm max} \sim N^\sigma,
\end{equation}
where $\sigma$ is a critical exponent of the transition. We have
\begin{equation}
\label{sigma-result}
\sigma=0.88\pm 0.06,
\end{equation}
which is close to $\sigma\!=\!1$. This implies that $S_2$ discontinuously changes against $b$. Thus, the result indicates that the model undergoes a first-order surface fluctuation transition between the smooth phase and the non-smooth phase, where "the smooth" phase is characterized by small $S_2/2N_B$, and "the non-smooth" phase is characterized by large $S_2/2N_B$. The dashed line in Fig.\ref{fig-5}(b) is drawn by fitting the smallest four data to Eq.(\ref{sigma-fitting}) and is consistent with a continuous transition when $N\!\leq\!9577$. This implies that the surface fluctuation transition is strongly influenced by the size effect.

\begin{figure}[hbt]
\centering
\includegraphics[width=9.0cm]{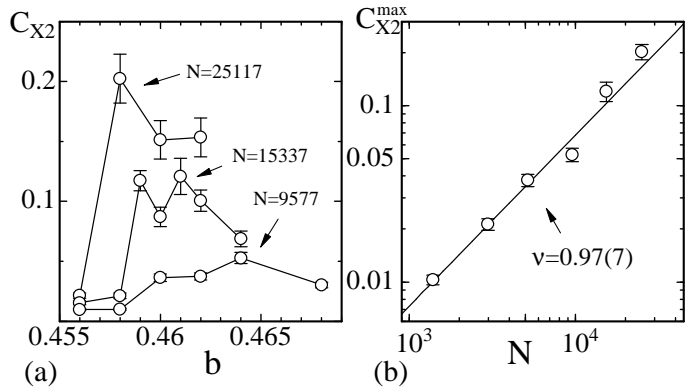}
\caption{The fluctuation $C_{X^2}$  vs. $b$, obtained on surfaces of (a) $N\!=\!25117$, $N\!=\!15337$, and $N\!=\!9577$, (b) a log-log plot of $C_{X^2}^{\rm max}$ against $N$ including the data obtained on the surfaces of $N\!=\!5167$, $N\!=\!2977$, and $N\!=\!1387$.}
\label{fig-6}
\end{figure}

The collapsing transition can be characterized by the quantity
\begin{equation}
\label{Spec-Heat-X2}
C_{X^2} = {1\over N} \langle \; \left( X^2 - \langle X^2 \rangle\right)^2 \; \rangle, 
\end{equation}
which is the variance (or the fluctuation) of $X^2$. Fig.\ref{fig-6}(a) shows $C_{X^2}$ against $b$. Anomalous peaks can be seen in the figure, and they indicate a collapsing transition, which separates the smooth phase from the collapsed phase, where "the smooth" phase is characterized by large $X^2$ and "the collapsed" phase is characterized by small $X^2$.   

%++++++++++++++++++++++++++++++++++
%++++++++++++++++++++++++++++++++++
Figure \ref{fig-6}(b) is a log-log plot of the peak value $C_{X^2}^{\rm max}$ against $N$. The straight line was drawn by fitting the data to
\begin{equation}
\label{nu-fitting}
C_{X^2}^{\rm max} \sim N^\nu,
\end{equation}
and then we have 
\begin{equation}
\label{nu-result}
\nu=0.97\pm0.07.
\end{equation}
Thus, we confirm that the collapsing transition is of first order. The exponent $\nu\!\simeq\!1$ in Eq.(\ref{nu-result}) implies that $X^2$ discontinuously changes against $b$. We note that the collapsing transition is not influenced by the size effect in contrast to the surface fluctuation transition.

%++++++++++++++++++++++++++++++++++
\begin{figure}[htb]
\centering
\includegraphics[width=4.5cm]{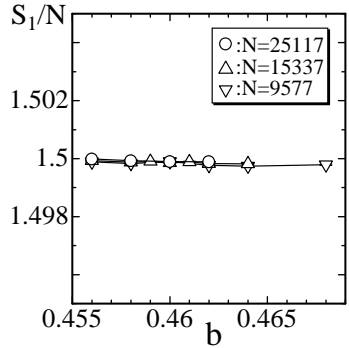}
\caption{The Gaussian bond potential $S_1/N$ vs. $b$ obtained on surfaces of $N\!=\!25117\sim 9577$. The expected relation $S_1/N\!=\!1.5$ is satisfied.  }
\label{fig-7}
\end{figure}
%++++++++++++++++++++++++++++++++++
The Gaussian bond potential $S_1/N$ are plotted in Fig.\ref{fig-7} in order to see that the expected relation $S_1/N\!=\!1.5$ is satisfied. The results clearly satisfy the relation, and therefore we consider that the canonical MC simulations were successfully performed.

%------------------------------------------
\section{Summary and conclusion}\label{Conclusions}
%------------------------------------------
We have studied an extrinsic curvature model on a disk in order to see whether or not the model undergoes a first-order transition between the smooth phase and the collapsed phase. Monte Carlo simulations were performed on lattices of size $N\!=\!25117\sim N\!=\!1387$. 

We found that two types of transitions occur at the same transition point; one is the surface fluctuation transition and the other is the collapsing transition. These transitions are well-known to occur in the same model on closed surfaces such as spheres and tori.

The surface fluctuation transition is strongly influenced by the size effect, and the simulation results obtained on the lattices of size less than $N\!=\!9577$ indicate that the model undergoes a continuous transition. However, the results obtained on the lattices larger than $N\!=\!9577$ indicate that the transition is discontinuous. The peak height of the specific heat scales according to $C_{S_2}^{\rm max} \sim N^\sigma$, and we have $\sigma\!=\!0.88\pm 0.06$, which is a sign of the discontinuous transition of surface fluctuation. 

The collapsing transition is not influenced by the size effect in contrast to the surface fluctuation transition. We have also obtained the fluctuation $C_{X^2}$ of $X^2$, and find the scaling property such that $C_{X^2}^{\rm max} \sim N^\nu$. Thus we have $\nu\!=\!0.97\pm0.07$, which is a sign of the discontinuous transition of surface collapsing phenomena.     

Combining the results in this paper and the previous ones obtained on closed surfaces \cite{Kownacki-Diep-PRE-2002,KOIB-PRE-2004-1,KOIB-PRE-2005,KOIB-NPB-2006,KOIB-PLA-2006-2}, we conclude that first-order transitions can be seen in extrinsic curvature surfaces, and the order of the transitions is independent of whether the surface is closed or open. 

%----------------------------------------------------------
%\vspace*{3mm}
%\noindent
%{\bf Acknowledgment}\\
%\vspace*{2mm}
%\par
%\section{References.}

%\vfill\eject

\end{document}